# The Transfer of Knowledge from Physics and Mathematics to Engineering Applications

**Roman Ya. Kezerashvili,**

*Physics Department, New York City College of Technology, CUNY, Brooklyn, NY 11201, USA*

**Candido Cabo,**

*Computer Systems Technology Department, New York City College of Technology, CUNY, Brooklyn, NY 11201, USA*

**Djafar K. Mynbaev**

*Electrical Engineering Technology Department, New York City College of Technology, CUNY, Brooklyn, NY 11201, USA*

## ABSTRACT

The objective of this paper is to describe a development of an innovative approach to enable students studying science, technology, engineering, and mathematics (STEM) to apply the concepts learned in physics and mathematics to engineering problem-solving challenges. Our goal is to facilitate students' transfer of knowledge from theory to engineering applications, thereby increasing student academic performance, retention, persistence, and graduation.

We have two primary and complementary foci that will facilitate students' ability to transfer knowledge of mathematics and physics to engineering applications: (1) To demonstrate the efficacy of using e-learning and e-teaching through Blackboard and Web-based communication systems as a means of providing more avenues of STEM learning; (2) to establish the laboratory as a primary learning tool in STEM at an early point in students' academic careers so that students have a taste of the excitement of science and engineering research. The hands-on learning experiences in the laboratory will be extended through the use of Blackboard system for interdisciplinary problem-solving activities. To achieve the goals we are creating a virtual community of students and faculty as a vehicle for promoting the transfer of knowledge using e-learning and e-teaching mechanisms.

**Keywords**: e-learning and e-teaching, virtual community, virtual class.

## 1. INTRODUCTION

There is growing concern that the United States is not preparing a sufficient number of students in the areas of science, technology, engineering, and mathematics (STEM) [1]. One of the effective methods to increase students' interest in STEM's area is to introduce e-learning and e-teaching technologies in the educational process [2]. Educational research has demonstrated the effectiveness of these technologies for interactive learning and lecturing techniques to encourage active learning [3-5]. For the other side it is important that student understand that knowledge of physics and mathematics are not abstract issues but have actual applications in engineering and technology. Therefore, the transfer of knowledge from physics and mathematics to engineering and technological applications becomes an essential factor to increase student academic performance, retention, persistence, and graduation in STEM.

The World Wide Web is becoming the learning tool of choice for many students. In order to use this environment effectively, faculty must create a student-centered community that is highly engaging and interactive. Interactivity comes from the interplay between laboratory experiences, lecture materials, online discussions, and assignments. In addition, the use of a variety of audio and video media appeals to a wider range of learning styles.

We are suggesting use of Blackboard and Web-based communications systems for transferring knowledge from physics and mathematics to engineering and technology for creating the virtual community of learner across disciplines. Our approach addresses a national problem in the teaching of engineering and technology subjects: students frequently have difficulty in applying foundational concepts in mathematics and physics to engineering and technology. By producing in our institution interactive interdisciplinary curricula that bridge mathematics, physics, and engineering subjects, we will not only improve retention and graduation of our STEM students but will make available, through a variety of dissemination mechanisms, new curricular products that can be used by other institutions with similar needs.

The article is organized as follows: Section 2 introduces the virtual community; in Section 3 we describe the architecture of the system, Sections 4, and 5 presented physics and engineering technologies modules, respectively. Conclusions follow in Section 6.

## 2. VIRTUAL COMMUNITY

To achieve our goals we suggest creating a virtual community of students and faculty as a vehicle for promoting the transfer of knowledge using e-learning and e-teaching mechanisms to extend the educational impact of real-time laboratories.

The virtual community is built using a Blackboard system and Web-based communication systems. Our approach utilizes a course management tool, Blackboard, - innovative ways to teach and learn, and to communicate and collaborate across disciplines. By putting courses online instructors of different disciplines, now strive to create a networked learning environment among different departments. At the same time, instructors face many challenges: increasing efficiency by doing more with less, addressing the needs of non-traditional students and finding more effective ways to measure student learning outcomes. Therefore, Blackboard becomes a primary tool for learning. Faculty members have the ability to control the pacing and sequencing of information to meet the cognitive needs of students. They can make new material available after students demonstrate mastery of previous material. They can also individualize the learning through resources available on the Internet and on networked CD-ROMs. Typical methods of teaching across disciplines can easily be adapted to the Blackboard system. For example, case-based situations and project-based learning can be designed as group work. Each group has their own discussion board, chat room, and e-mail capabilities. A team leader can be assigned to the group to direct the group in a manner similar to a real-life scenario. These activities will complement the laboratory explorations that the real-time laboratories will support. Experts can also be introduced in the online environment to add their perspectives and describe how they applied skills and concepts to the specific discipline.

Within the virtual community, virtual classes are creating, which allow the instructors of physics, mathematics, and engineering to communicate with each other and work with the same students simultaneously. Key concepts, applications and techniques can be exchanged more readily in any discipline. Feedback from the engineering professors helps the instructors in physics and mathematics identify specific topics and their applications that are emphasized in the technology and engineering courses. Therefore, engineering instructors will not need to review as much. The Web-based system supports the development of a community of learners, who can share ideas and best practices in an application-oriented environment. Students appreciate the learning of physics and mathematics when the knowledge is transferred. Students studying in a virtual classroom will be monitored in their major as well as other departments, thus creating a community that cares about their academic success.

In virtual classes students of engineering technology will study physics, mathematics, and engineering technology subjects while being enrolled in regular academic classes. This step will result in increasing students' interest in studying physics and mathematics courses, as they will immediately see the necessity and the usefulness of these subjects in their majors. This will also increase students' interest in studying engineering technology subjects, as they will immediately see the scientific foundation of the specific devices and processes used in technology. This mutual enrichment of science and technology study will practically result in the increase of retention and graduation rate of the enrolled students and will attract more students to engineering technology programs.

## 3. ARCHITECTURE OF THE SYSTEM

Our approach is based on Blackboard e-Education platform and Website communication as vehicles for instructional delivery. The website is accessible to all: students taking particular courses in all discipline, instructors teaching these courses, as well as to anyone who has access to WWW. From this point of view the system is open. The website menu links to courses of physics, mathematics, computer information system and electrical engineering and telecommunication technology. The courses have cross-references. Let's say a student studying the application of Ohm's law or Kirchhoff's rules in a course of electrical circuit and needs to solve a linear or quadratic equation. When words "Ohm's law" or "Kirchhoff's rules", "linear equation" or "quadratic equation" appear in this course by clicking on these words, the system will link to the corresponding lecture material in physics or mathematics, where the student will find the detailed explanation of these concepts, and methods of solutions equations and thus, refresh these materials. From the other side, by studying let's say "Total Internal Reflection" in a physics course the student can link to application of this concept in fiber optics communications and therefore, students immediately realize that "Total internal Reflection" concept has real application in his/her major. The Website is also linked to WWW. These links are established by instructors and extend lecture materials to the best available and suitable information on the WWW, related to the subject. Actually, just by using the Website students are only learning their own Web-based teaching materials without any interaction with other students and instructor. This learning experience must be definitely worse than that of traditional classroom instruction. Introduction the Blackboard system gets closer to the

235

ideal state of individualized instruction and interactive learning and teaching.

Each course of physics, mathematics, computer information system and electrical engineering and telecommunication technology from the Website is linked to corresponding course at the Blackboard. By logging into Blackboard the student becomes participant of the virtual classroom. The Participant Area of the Blackboard system displays the names of all of the participants in the Chat session and allows the Instructor to manage their participation. This area also displays which participants have requested to speak and which have been recognized. The Content Map, similar to the Course Map, is available within Virtual Classroom sessions. The Communication area allows students and instructors send email, access Discussion Boards. Using the Communication area instructor can bring top publisher content into e-Learning. E-links established by instructors allow students of the virtual class to visit related materials on the World Wide Web and log into the online library. The Blackboard system is individualized and available only for students which are registered in virtual class. Fig. 1 presents the architecture of the system. In this figure CST and EET stand for Computer Systems Technology and Electrical Engineering Technology, respectively.

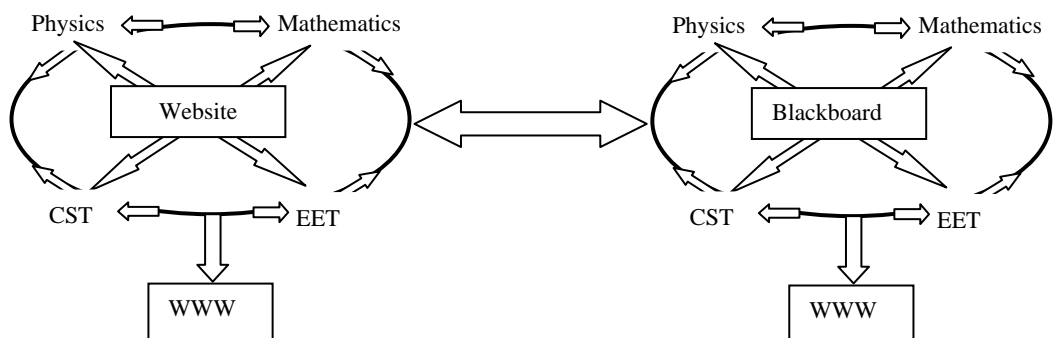

Fig.1. Architecture of the e-learning and e-teaching system.

## 4. PHYSICS MODULE

The traditional didactic college physics lecture usually is a teacher-centered lecturing environment in which the instructor does all the talking and some times supports the explanation by demonstrations. This ancient method still remains the dominant pedagogical practice. This traditional lecture format gives emphasis to the transfer of knowledge from an instructor,-the proprietor of knowledge and from real-time experiment,-accumulator of knowledge, to student,- the beneficiary of knowledge. Thus, in teaching physics always three components are involved: instructor, as the holder of information, student, as the recipient of information, experiment, as the confirmation and visualization of knowledge. The traditional lecture format fails to provide the interactions between these components, and as a result gives little opportunities for majorities of students to be involved in discussions and asking questions. But this is a key for development of critical thinking and problem-solving skills.

Interactive engagement is now recognized as an important technique in introductory physics courses. Educational research has shown that traditional lecturing, in which students listen passively, results in surprisingly little real learning. Instead, teaching styles in which students actively engage in discussing questions with each other, actively work out answers for themselves, and interact with the instructor, are measurably more effective in imparting real understanding.

Physics cannot be taught only using the book and blackboard and asking students to memorize rules, formulas and laws. One of the important parts of teaching physics is real-time experimental demonstration which visualizes the laws of nature as well as laboratory experiments which students conduct during laboratory sessions. The laboratory and experimental demonstration should be establish as a primary learning tool in STEM at an early point in students' academic careers so that students have a taste of the excitement of science and engineering research. Therefore, it is necessary to promote teaching and learning resources which supports "Instructor-Student-Experiment" interactive engagement.

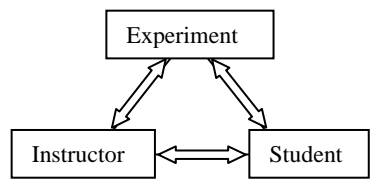

Fig.2. Interactions of the components "Instructor-Student-Experiment"

E-learning and e-teaching can bind this "triangle" through the interaction of these components as shown in Fig. 2.

Physics module of our system, which contains two courses of algebra-based general physics, reflected all of the above mentioned features and included several teaching resources aimed to promote comprehension of the physics laws: class lecture, demonstration experiments, laboratory experiments, e-learning material, problem-solving sessions. Class lecture presents one of the most important principles for every physics course, - *concepts first*. Conceptual understanding is the focus through the explanations, examples and media demonstrations of the experiments and is presented on the Website and on the Blackboard. E-learning material provides problem solving examples, and problem-solving session provides to students through the interactive system between student and instructor **"Physics Tools"** [1], which is a Problem-Based Learning and Problem-Solving tool, given on the website. The architecture of our system is implemented primary in HTML with embedded Java scripts and Java applets. The system allows students to check the analytical solution of the problem, numerical value and units, provides the hints and the detail solution of the problem by a request, evaluates a student's work and gives the correct answer by a request. Physics Tools module consists from two shells. One shell is intended for students and other is intended for an instructor. Physics Tools is designed to help students to understand main principles and concepts of physics through visualization experimental phenomena and performing virtual experiments and developing problem-solving skills. The objectives of Physics Tools are to improve the success rate of students enrolled in algebra and calculus-based physics classes and to develop critical thinking skills. This system provides students an outstanding training and preparation basis - exactly what they need in order to score well on typical physics exams.

The physics module provides a demonstration of real physics experiments from mechanics, liquids, oscillations and waves, electricity, magnetism and optics. Each experiment stand-alone and illustrate a particular principle of physics and integrated in the lecture materials. In developing this module we are using existing 25 DVDs of the video encyclopedia of physics demonstrations [6], some demonstration suggested by us [7,8] and digital taping them as well as making the links to existing Physics Lecture Demonstrations on Internet [9]. Most video experiment have parallel soundtracks and detail descriptions. All these will permit students to listen and read the explanation of the physics phenomena. Student may repeat the experiment few times and concentrate on learning and understanding the physical processes. All these allow for interactive participation by students for studying and understanding physics concepts and principles and their applications in engineering.

## 5. ENGINEERING TECHNOLOGY MODULES

### 5.1 COMPUTER SYSTEMS MODULE

The pedagogical approach taken to teach courses in Computer Systems is similar to the interactive engagement method described earlier in the physics module section. This methodology that we refer to as "hands-on + theory" is equivalent to the lecture (theory) and experiment (hands-on) combination described earlier. We found that the hands-on and theory combination is synergistic. An understanding of the theoretical concepts helps students in the solution of the computer laboratory exercises. At the same time, working on the solution of a specific problem in the laboratory brings out a deeper understanding of the theoretical concepts explained in the lectures. The understanding of the theory remains shallow without the hands-on computer labs, and the hands-on exercises without the theory are little more that random trial and error attempts at solving a problem.

Furthermore, we are trying to blur the line between lecture and laboratory. Lectures in computer systems courses are conducted in computer laboratories where each student has access to a computer. As the instructor explains a theoretical concept using the blackboard or a PowerPoint presentation, s/he can illustrate the concept with an application using the instructor's computer. The instructor's monitor is projected in a screen for all students to see. Students can them reproduce what the instructor does in their own computer. We found that this approach of practical illustration of theoretical concepts has helped students, in a wide range of computer systems courses (from computer programming to database design and networking) to develop a deeper understanding of the concepts and to develop problem-solving skills that they can use later in projects and laboratory exercises.

Another important component of student learning is the availability of educational tools outside the classroom and computer laboratories. Obviously the amount of time spent in the classroom or laboratory is not enough for students to assimilate the theoretical concepts learned in class and to develop the adequate practical skills. Therefore student access to instructional material through Blackboard and software applications (compilers, database management systems, network analyzers) used in the laboratories is crucial. We are currently developing the architecture for external access to all our software applications (residing in the departmental servers) from outside the college. Despite the fact that there are several open laboratories in the college where the students can work outside the classrooms, access from outside the college is very important for students in a commuter college like City Tech.

The development of the Virtual Community described here helps to put together all the teaching/learning practices that we found help our students to learn. But this Virtual Community is much more; it helps the transfer of knowledge from physics and mathematics to computer systems. Many basic concepts in computer systems come from mathematics and physics. However, students have a hard time at identifying that connection. This Virtual Community is an attempt to make the connection more obvious by using the syntax of the Web: hypertext. At the present time we are incorporating a sequence of three upper level courses on networking in the Virtual Community. All the lecture material is incorporated in the web site. Most importantly, the material incorporates hypertext links to relate concepts taught in the networking courses to related concepts in mathematics and physics. In addition to the lectures, we plan to incorporate laboratories in the Virtual Community web site. Those laboratories could also be linked to related laboratories or problems in mathematics and physics. Also important is the use of simulation tools like the network visualizer RouterSim and the virtual-machine simulator VMware. Those simulation tools allow students to reproduce in the virtual environment of the internet the real environment of the physical computer laboratory. Even though simulation tools do not substitute the practical exercises in the physical laboratory, they allow students to test possible solutions to real problems that can later be tested in the real laboratory.

### 5.2 ELECTRICAL ENGINEERING TECHNOLOGY MODULE

Electrical engineering technology courses are taught in interactive mode based on combination of theory and experiments and the electrical engineering technology module comprises, at this stage, three courses: Network Analysis I, Network Analysis II and Electronics. These courses include analysis of dc and ac circuits and basic electronics; they are, in essence, the first courses in major that electrical engineering technology students meet in their study. The courses of Network Analysis I and II combine both theory and laboratory segments; course of Electronics is a theory course only and laboratory exercises for this course are conducted as a separate course.

Engineering technology programs always rely on hands-on experiments and our circuit analysis courses can serve as good examples of such approach. During every week of study, the students learn a specific topic in theory and confirm this concept via experiment. For example, students study the Ohm's law in a traditional lecture and then confirm its validity by measuring a voltage-current relationship in the laboratory. In our relevant module we provide students with derivation of the basic formula and discussion of this formula in various formats. The focus of this study is the application side of the Ohm's law; in other words, we concentrate at changing current flowing through a resistor caused by changing applied voltage. In a corresponding experiment, students do actual measurements of current while changing voltage.

With this study, students grasp the concept of the Ohm's law. To introduce this law (and other topics for that matter) from different standpoint, we use computer-based experiments. Students build a circuit at a computer screen and run the simulation program. The simulated circuit includes the measuring instruments and the students can actually read the measurements from a voltmeter and an ammeter displays. Such virtual experiments enhance students learning: when they turn later to the real-time experiment, they feel more comfortable working with tangible equipment since they have already an idea how this circuit works and what results they should expect. In addition, students develop the skills needed for their future work because such an approach is typical in an industrial environment. What's more, students learn how to practically use such a popular industrial simulation tool as MultiSim.

Electrical engineering technology courses include, of course, sets of problems and solutions. In our example of the Ohms' law, students exercise in computing values of current while varying applied voltage with a given resistance. They also learn through problem-solution approach how value of a resistor changes the slope of a voltage-current graph. This approach enhances the students learning through using practical examples. It also helps students to understand the range of values of resistor, voltages and currents they will encounter in their future workplaces.

All courses developed for this module are presented in Blackboard, which allows the instructors use all advantages of this e-learning and e-teaching tool, as described in the previous sections.

While concentrating at application side of the topic in study, our students must, of course, understand the main physical and mathematical concepts supporting these applications. In our example of the Ohm's law, students have to understand why current is directly proportional to voltage and why the value of a resistor changes the slope of the corresponding graph. Here the direct connection with physics module comes into the play. By clicking at the proper physics module, students can refresh their memory on physical mechanisms that determine such phenomena as current, voltage and resistance. Deep understanding the physics behind these phenomena provides students deep –knowledge of the qualitative and quantitative relationship among them. Similarly, solving the problems related to the Ohm's law, students can link to the appropriate mathematics module and refresh their knowledge of linear equations. Building the system presented in Fig. 1, we make these connections easy to use and always available. In our system all necessary knowledge is literally "a click away" for every student involved in this study.

### 6. CONCLUSIONS

Our approach proves the necessity of the innovative computer-based approaches to teaching a new generation of highly-technical students. It breaks down the traditional barriers among different disciplines, and changes the teaching culture of academic institutions. All these will bring the main result: better preparation of our students to their future professional work and to their future lives.

This approach, on one hand, will enhance the computer-based teaching in the mathematics and physics programs, which will result in a greater involvement of the mathematics and science faculty members in the e-teaching process. On the other hand, it will enhance the e-learning and e-teaching approach, which will result in greater involvement of the students into the computer-based learning process.

The use of e-learning and e-teaching approach based on the Blackboard and Website communication system allowed greater student-instructor and student-student interaction.

*Acknowledgement:* This work is supported by US Department of Education under the Grant P120A060052.